# Damage threshold in pre-heated materials exposed to intense X-rays

Nikita Medvedev[1,2,*], Zuzana Kuglerová[1,3,4], Mikako Makita[5], Jaromír Chalupský[1], Libor Juha[1,6]

[1]*Department of Radiation and Chemical Physics, Institute of Physics, Czech Academy of Sciences, Na Slovance 1999/2, 182 21 Prague 8, Czech Republic*

[2]*Laser Plasma Department, Institute of Plasma Physics, Czech Academy of Sciences, Za Slovankou 3, 182 00 Prague 8, Czech Republic*

[3]*Department of Surface and Plasma Science, Faculty of Mathematics and Physics, Charles University in Prague, V Holešovičkách 2, 180 00 Prague 8, Czech Republic*

[4]*Deutsches Elektronen-Synchrotron DESY, Notkestraße 85, D-22607 Hamburg, Germany*

[5]*European XFEL GmbH, Holzkoppel 4, D-22869 Schenefeld, Germany*

[6]*Engineering Research Center for Extreme Ultraviolet (EUV) Science and Technology, Colorado State University, Fort Collins, Colorado 80523-1320, USA*

[*] Corresponding author: ORCID: 0000-0003-0491-1090; Email: nikita.medvedev@fzu.cz

## Abstract

Materials exposed to ultrashort intense x-ray irradiation may experience various damaging conditions depending on the *in-situ* temperature. A pre-heated target exposed to intense x-rays plays a crucial role in numerous systems of physical-technical importance, ranging from the heavily-, and repeatedly radiation-loaded optics at x-ray free-electron laser facilities, to the first wall of prospective inertial fusion reactors. We study theoretically the damage threshold dependence on the temperature in different classes of materials: an insulator (diamond), a semiconductor (silicon), a metal (tungsten), and an organic polymer (PMMA). The numerical techniques used here enable us to trace the evolution of both, an electronic state and atomic dynamics of the materials. It includes damage mechanisms such as thermal damage (induced by an increase of the atomic temperature due to energy transfer from x-ray-excited electrons) and nonthermal phase transitions (induced by changes in the interatomic potential due to excitation of electrons). We demonstrate that in the pre-heated materials, typically, the thermal damage threshold stays the same or lowers with the increase of the *in-situ* temperature, whereas nonthermal damage thresholds may be lowered or raised, depending on the particular material and specifics of the damage kinetics.

## 1. Introduction

In the extreme ultraviolet (XUV)/x-ray irradiation experiments, typical experiments are performed with solid targets initially kept at room temperature due to the simplicity of sample preparation and operation. There are reasons, however, to study the consequences of an initial temperature well above room temperature for materials under XUV/x-ray laser interactions. Such studies can help in understanding the behavior of the following systems:

(a) Optical elements (e.g., mirrors, lenses, filters, windows) utilized for guiding, tailoring, and focusing XUV/x-ray laser beams are thermally loaded, especially those used at high-repetition-rate sources. The level of radiation and thus the thermal load determines the survival of such optics most typically with (kHz – MHz repetition rate) XUV/x-ray free-electron laser pulses [1–3]. For such high-repetition cases, each subsequent x-ray pulse arrives at an optical component shortly after it was irradiated (*i.e.*, heated up) by the previous pulse – the scenario, for instance, immediately raises concern for optics such as focusing optics, monochromators, and split-and-delay units [4]. A detailed study of radiation damage to pre-heated samples should also shed light on the action of short XUV/x-ray laser pulses delivered by plasma-based sources [5,6].

(b)  Material removal process enhanced by XUV/x-ray induced thermal effects (especially, the desorption-like processes [7]) can be utilized for an efficient nano/micro-structuring of technically important materials. Such direct processing can be envisioned to extend the EUV Lithography (EUVL) which currently dominates the production of integrated circuits. The thermal enhancement by XUV/x-ray-initiated erosion of solids can also be utilized in analytical imaging techniques [8], benefiting from a higher yield of liberated molecular ions achieved at elevated temperatures.

(c)  The first walls of inertial confinement fusion (ICF) reactors are repeatedly heated when being damaged by radiation and particle emissions from fusion plasmas (see for example Refs. [9,10]). It has been recognized that an elevated initial temperature can either enhance or reduce the radiation damage depending on the irradiation intensity, thermal conditions, and materials properties [10].

The standard assumption is that preheating of materials should lower the damage threshold proportionally to the heat deposited in the sample. This hypothesis, however, has not yet been validated. Moreover, the consideration of the total heat deposited in the material as the defining factor of the material response comes from considering only the thermal damage, whereas nonthermal effects may not have such a correlation with preheating. The synergy of the nonthermal damage with preheating has not been studied before. Since various types of materials experience different nonthermal mechanisms under irradiation [11], it is important to consider their response to irradiation.

For this study, we chose three representatives of elemental solids: diamond, silicon, and tungsten, and one example of molecular solids: organic polymer PMMA – Poly(methyl methacrylate). Diamond, silicon, and PMMA play a crucial role in the XUV/x-ray instrumentation (see for example Refs. [1,2,12–16] and references therein), while tungsten and its alloys are widely considered the first-choice material of first walls in prospective ICF reactors (see Refs. [9,10] and references therein). We apply computer simulations to study how x-ray damage thresholds in these materials depend on the initial temperature at which the sample is pre-heated.

## 2. Computer simulation techniques

Intense XUV/x-ray irradiation of matter triggers a sequence of processes, ultimately leading to observable material damage [17]. It starts with the excitation of electrons *via* x-ray photon absorption, which is the dominant channel of photon interaction with matter in the XUV/x-ray regime [18]. It transiently brings the electronic system out of equilibrium, where photo-excited electrons produce secondary cascades, exciting new electrons from the valence band or core shells to the conduction band of the material. At the same time, created core-shell holes decay via the Auger channel, also exciting secondary electrons. For the considered here photon energies and materials, the radiative decay channel plays only a minor role in core-hole decays [19]. Secondary electron cascades induced by XUV/x-ray irradiation typically end within a few tens of femtoseconds, when all electrons lose their energy below a certain threshold and cannot ionize new electrons [20,21].

Electrons promoted to the conduction band affect the atomic system of the irradiated material in two ways: (i) they exchange kinetic energy with atoms via scattering (so-called, electron-phonon coupling), and (ii) excitation of electrons modifies the interatomic potential (so-called, nonthermal effects) [22]. The channel (i) increases the kinetic energy of ions of the target, which upon overcoming the melting temperature of the material may induce a phase transition. In contrast, channel (ii) changes the forces acting on the atoms, which at high levels of excitation may destabilize the lattice and induce an ultrafast phase transition known as nonthermal melting [23,24].

To describe the observable material damage, all the above-mentioned effects need to be included in an appropriate model. To this end, we apply the hybrid simulation tool XTANT-3 [25]. This code combines a few simulation techniques into a unified approach. All the details of the code may be found in the Refs. [25,26], here we only briefly describe the essential points of the simulation.

XTANT-3 traces the electronic excitation employing the transport Monte Carlo (MC) method [27]. The MC model traces the photoabsorption using cross-sections available from the EPICS2017 database [19]. The Auger-decay times of core holes are sampled using the characteristic times from the same database. The electronic scattering cross sections applied are the binary-encounter Bethe for inelastic scattering (impact ionization) [28], and the modified Mott cross section with Molier screening parameter for elastic scattering [29]. In the present work, we will not be focusing on the non-equilibrium stage, and the electronic excitation only serves as a means to heat the low-energy fraction of the electronic ensemble, populating the valence and bottom of the conduction band.

The electronic band structure is calculated using the transferrable tight binding (TB) method [25]. The TB Hamiltonian enables the calculation of the evolution of the electronic energy levels (molecular orbitals, or band structure), the electronic wave functions, and the potential energy surface (forces) of atoms. Excitation of electrons directly affects all these quantities in the TB formalism, thereby allowing to take into account nonthermal effects. We applied the following TB parameterizations in this study: to model diamond, we used the parameters from Ref. [30] which relies on an orthogonal basis set of the linear combination of atomic orbitals; for silicon, we applied the same kind of parameterization from Ref. [31]; for tungsten, the non-orthogonal NRL TB parameters are used [32] (with added short-range repulsion stabilizing the system at high temperatures [26]); and for PMMA, the matsci-0-3 DFTB parameterization is employed [33].

To trace the kinetic energy exchange between the electrons of the valence and conduction band with atoms, XTANT-3 uses Boltzmann collision integrals (BCI) formalism [26]. The electron-ion collision integral depends on the transient matrix elements of the electron-ion interaction, and the electron populations (distribution function). Utilizing the so-called "bump-on-hot-tail" transient distribution of electrons after XUV/x-ray irradiation [34,35], the low-energy fraction of electrons which is at near-equilibrium conditions is described with Fermi-Dirac distribution function (whereas the nonequilibrium tail is traced in the above-mentioned MC module). The matrix elements are calculated with help of the transient wave functions, available from the TB Hamiltonian. In this way, the BCI formalism allows capturing the energy exchange between electrons and atoms in response to the atomic motion (such as, *e.g.*, atomic vibrations, phonons).

The atoms are traced with the molecular dynamics (MD) simulation. The interatomic potential, evolving together with the changes in the state of the electronic system, is provided by the transient TB Hamiltonian. Additionally, the energy transferred from electrons (both, *via* elastic scattering of fast electrons traced within MC, and slow electrons modeled with the BCI), is fed to atoms *via* the velocity scaling algorithm at each time-step of the simulation [26]. We apply the 4$^{th}$ order Matryna-Tuckerman algorithm to propagate atomic coordinates with the timestep of 0.1 fs [36]. Over two hundred atoms are used in each simulation box. The simulations are initialized with atoms in their equilibrium positions and random velocities assigned according to the Maxwellian distribution. Before the arrival of an XUV/x-ray pulse, the system is allowed to equilibrate for a few hundred femtoseconds [26].

All the models involved are executed simultaneously and are interconnected within XTANT-3. That creates a hybrid simulation capable of tracing all the important transient processes, resulting in material damage after ultrafast XUV/x-ray irradiation, as has been validated by comparisons with experimental data, see e.g. Refs. [25,26,37,38].

To identify damage thresholds, a series of simulations are performed in each material at various initial temperatures reached by a pre-heating of the sample. For each temperature, we identify a threshold dose (deposited energy density), at which damage onsets. This allows us to construct the damage threshold as a function of the initial atomic temperature. Below we discuss specifics of the damage in each of the studied materials and present the results of simulations.

In addition, we performed a separate set of simulations for diamond and silicon using hybrid code TREKIS-4 [39]. This code is based on a combination of the transport MC method (similar to the MC module described above for XTANT-3), and classical MD simulation. We employed Stillinger-Weber potential [40,41], because it reproduces the melting point of silicon better than most of the other potentials available [42]. Having an empirical potential, TREKIS-4 does not model directly the nonthermal effects such as an influence of the electronically excited state on the interatomic potential, but approximates it as an increase of the kinetic energy of atoms in response to formation of the electron-hole pairs [39]. Thus, TREKIS-4 is best suited to model thermal effects, which allows us to cross-check XTANT-3 calculated damage threshold with this independent tool and additionally validate our conclusions.

## 3. Results and discussion
### 3.1 Diamond

To evaluate the damage threshold in diamond under XUV/x-ray irradiation, we use an NPH (isoenthalpic-isobaric) ensemble, allowing for the simulation box to expand, adjusting the geometry to the atomic structure according to the Parrinello-Rahman method [43]. It is important when we model phase transitions between material phases with different structures and densities [27]. It assumes that the irradiated sample in the experiment would be a finite-size target able to expand, such as a thin layer of diamond or its near-surface regions. We perform a series of simulations with XTANT-3 with various irradiation doses with a step of 0.1 eV/atom to identify at which dose material phase transition takes place. The phase transition may be identified directly *via* atomic snapshots, see the example in Figure 1, where diamond graphitizes within ~200 fs within the NPH ensemble. VMD software is used for visualization of the atomic snapshots [44].

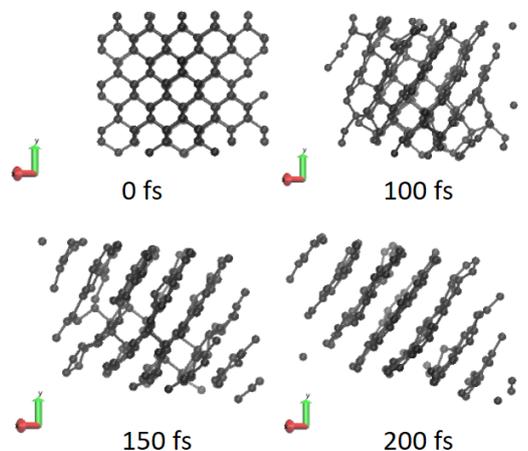

*Figure 1. Snapshots of diamond irradiated with 1.5 eV/atom deposited dose, simulated with NPH ensemble in XTANT-3 code.*

The calculated damage threshold for various atomic temperatures is shown in Figure 2. The shown temperature range is limited by the equilibrium graphitization threshold temperature (~1200 °C [45]; since equilibrium graphitization requires macroscopic timescales, at ultrashort simulation timescales presented here, a diamond can transiently withstand somewhat higher temperatures).

We see an overall tendency for the damage threshold to decrease with the increase of the irradiation temperature. Moreover, the decrease in the damage threshold dose cannot be explained solely by the energy present in the diamond due to pre-heating. If we add the energy associated with the pre-heating, which increases the kinetic energy content in the sample, the resulting dose is lower than the room-temperature damage threshold, see the dashed line in Figure 2. The energy content provided via the pre-heating and the irradiation does not add up to a constant (straight line). This indicates that some non-linear effects play an important role, such as anharmonicity of the interatomic potential. Atoms, already oscillating at high temperatures, may overcome the potential barriers more easily as the barriers are lowered due to the electronic excitation [11]. Such a synergy of thermal and nonthermal effects lowers the damage threshold in XUV/x-ray irradiated diamond with the increase of the *in-situ* temperature.

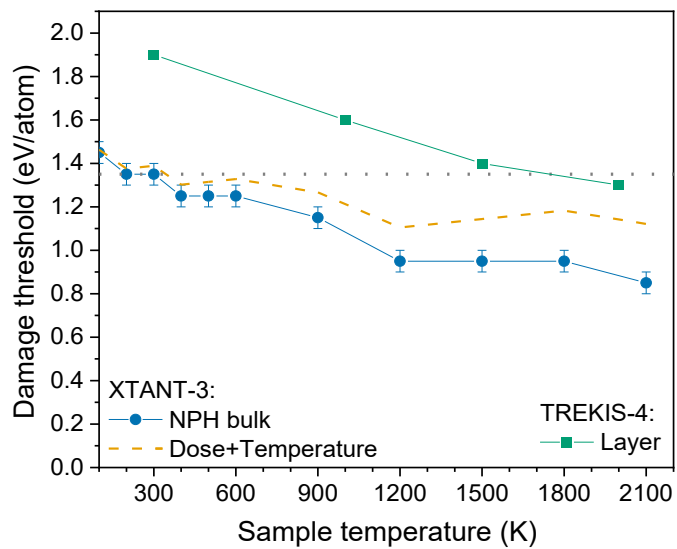

*Figure 2. Damage threshold in diamond vs. irradiation temperature, simulated within NPH ensemble with XTANT-3 code (blue circles), in diamond layer with free boundaries along Z-axis (black triangles), and the same layer simulated with TREKIS-4 code (green squares). The dashed line is the calculated threshold with added kinetic energy associated with the pre-heating of the material (for NPH simulation). The dotted horizontal line marks the room-temperature threshold for comparison.*

Let us mention that the simulation with an NPH ensemble, allowing the simulation box to adjust the volume (density) in response to a phase change, assumes that the target is unconstrained. This applies to near-surface regions of a material, to finite-size samples, or, perhaps, to polycrystalline samples. In the bulk, the behavior is different, since a material that cannot expand does not form clear graphite but a mixture of graphite and diamond (or amorphizes at even higher doses above the threshold) [46].

We also note that the damage threshold at the room temperature in Figure 2 is higher than the earlier reported value of ~0.7 eV/atom from Ref. [27], and ~1 eV/atom reported in the more recent Ref. [46]. The main reason for this difference is that the electron-phonon coupling was accounted for in the present simulation, whereas the earlier works were performed within the Born-Oppenheimer approximation which excludes non-adiabatic (electron-phonon) coupling. Including electron-phonon coupling here reduces the electronic temperature, which, in turn, affects the interatomic potential.

The potential formed at cooled electronic temperature is closer to the unexcited potential. Therefore, it requires higher deposited doses to reach the damage threshold, compensating for the cooling of electrons due to coupling.

The simulations of the thin layer of diamond with free open surfaces, which are free from the assumptions of the Parrinello-Rahman NPH ensemble [47], demonstrated the same behavior, however with a lower absolute value of the damage threshold of ~0.7 eV/atom at the room temperature, reducing down to ~0.4 eV/atom at 2000 K. An additional simulation with the TREKIS-4 code also showed the same qualitative behavior of the damage threshold reduction with increase of the *in-situ* temperature (see Figure 2). The TREKIS-4 predicted threshold is higher due to the usage of the classical MD potential, softening of which due to electronic excitation is not accounted for. Nevertheless, the fact that the qualitative behavior of the damage threshold with the rise of the atomic temperature is independent of the simulation method and the numerical tool employed validates the conclusions.

## 3.2 Silicon

To study the response of silicon to irradiation, we use the same methodology as in the last section. As was previously shown, silicon under irradiation can undergo two different phase transitions, depending on the deposited dose: thermal transition into a transient low-density liquid phase (LDL), or an interplay of nonthermal and thermal transition into a high-density liquid (HDL) phase [48]. The LDL phase is unstable and, presumably, should either transition into the HDL phase or resolidify. This is expected to take place at long timescales, nanoseconds or longer, which cannot be reached with the present simulation. We trace the transient damage that takes place within 2 ps of the simulation time and study the dependencies of the LDL and HDL damage formation thresholds.

As known from the previous work [48], we can identify the LDL phase formation by the band gap collapse, whereas the HDL phase can be identified by shrinkage of the simulation box (reduction of volume) [48]. In a sequence of simulations with various deposited doses, we identify the thresholds where the formation of new phases starts for each studied starting temperature.

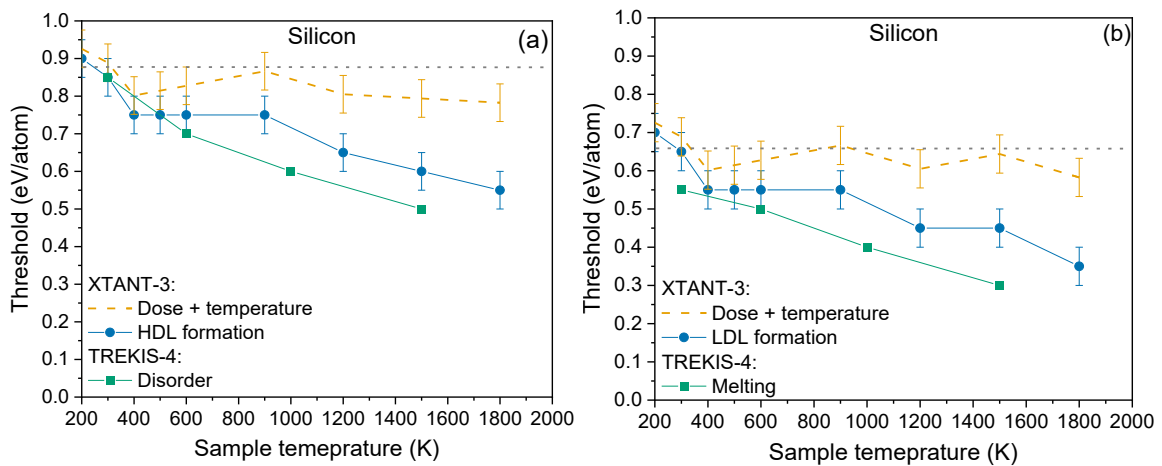

*Figure 3. Damage thresholds in silicon vs. irradiation temperature, simulated within NPH ensemble with XTANT-3 code. Thresholds for ultrafast transitions into (a) HDL and (b) LDL phases are shown. Points are calculated damage threshold, connected by lines to guide the eye: blue circles are calculated with XTANT-3, green squares are those calculated with TREKIS-4. The dashed line is the calculated threshold with added kinetic energy associated with the pre-heating of the material. Dotted horizontal lines mark the room-temperature thresholds for comparison.*

The results are shown in Figure 3. The overall tendencies are the same as in the diamond reported above. With the increase of the target temperature from 200 K to 1800 K, the threshold for ultrafast HDL phase formation reduced from 0.9 eV/atom to ~0.6 eV/atom, *i.e.* by 33%. This reduction is only partially accounted for by including the energy of the pre-heating (see the dashed line in Figure 3a).

In contrast, the LDL phase formation threshold can be fully explained by the additional energy provided *via* pre-heating, see Figure 3b. Here, the dashed line coincides with the room-temperature threshold (dotted horizontal line) within the error bars. The thermal phase transition into the transient LDL phase is accompanied by only minor modifications of the interatomic potential, and thus the atoms require the same amount of energy to overcome the potential barriers in each case.

In addition, Figure 3 presents the results of simulation with TREKIS-4 for identical conditions as those modeled with XTANT-3 reported above. As TREKIS-4 does not trace the electronic structure evolution [39], we instead identify the two damage thresholds by the onset of melting (the appearance of first defects, corresponding to LDL phase) and as the complete disorder (corresponding to the HDL phase). Since XTANT-3 predicted that the major contribution to the damage in irradiated silicon is from the thermal damage, a comparison with the classical MD simulation in TREKIS-4 is justified.

Figure 3 shows that TREKIS-4 predicts very similar behavior of the damage thresholds with increase of the *in-situ* target temperature. The TREKIS-4 simulation results are reasonably close to those of XTANT-3. A slightly lower absolute values calculated with TREKIS-4 seem to stem from the application of the empirical interatomic potential, in contrast to the TB-calculated one in XTANT-3. Thus, we again conclude that the qualitative behavior reported is independent of the simulation tool and should be observable experimentally.

### 3.3 Tungsten

To trace explicitly the onset of the melting and ablation process (mass removal of material from the surface of the irradiated target) in tungsten at different irradiation temperatures, we use a thin layer of material in the simulation (~2 nm). The simulated tungsten has free surfaces along the Z axis and periodic boundary conditions along X and Y [49]. The melting process in W is mainly thermal, as nonthermal effects in metals contribute to electronic pressure, possible expansion, and ablation, which take place at higher electronic temperatures [49]. At the thermal melting threshold, the damage starts from the surface. The surface atoms start to disorder at lower doses and shorter timescales than those inside the bulk. Then, the melting front propagates into the depth of W (this process was investigated previously in the literature theoretically in [50] and experimentally in [51]).

To identify the thermal melting in a metal sample, we used the following procedure: first, we slowly homogeneously heat the electronic system in the simulation box (with a laser pulse of 5 ps FWHM duration), tracing the atomic response to it. The onset of melting can be identified by a change of the slope in the atomic mean displacement, see an example in Figure 4. The kink occurs at the mean displacement of ~0.25 of the nearest neighbor distance ($a$=2.7 Å in W). This threshold can be used as a criterion for the onset of melting, similarly to the Lindemann criterion [52]. We see in Figure 4 that at the time instant of the melting onset, the average atomic temperature is ~3400 K, which is close to the experimental melting point of tungsten (3695 K). The calculated melting temperature is slightly lower than the experimental one, since the melting in our ultrathin sample starts from the surface, in which case the melting temperature is expected to be lower than the one in the bulk. We conclude that the criterion of the mean atomic displacement of ~0.25$a$ is consistent and may be used in further simulations. Knowing the time instant of the melting onset, we can identify the deposited dose at this point according to the profile of the pulse used in the simulation. This value will be used as the definition of the melting threshold dose.

We then performed a series of analogous simulations with various starting atomic temperatures, from room temperature up to the melting point of tungsten, and identified the melting threshold doses. The resulting melting threshold in tungsten is shown in Figure 5.

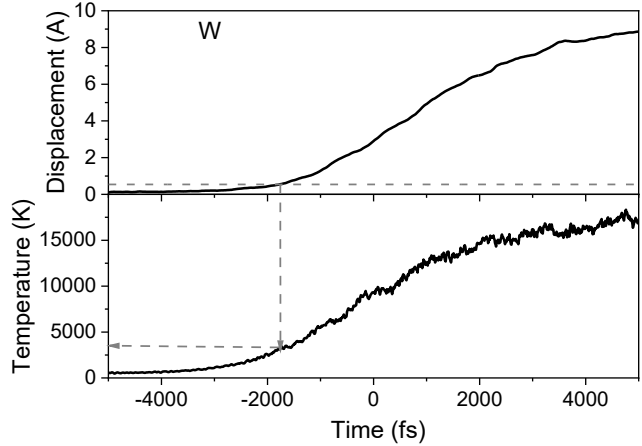

*Figure 4. Top panel: mean atomic displacement in a layer of tungsten, heated with a pulse of 5 ps FWHM duration, 10 eV/atom total dose. The grey line shows the onset of melting. Bottom panel: the atomic temperature in the same simulation. The grey arrows point to the temperature at the melting onset.*

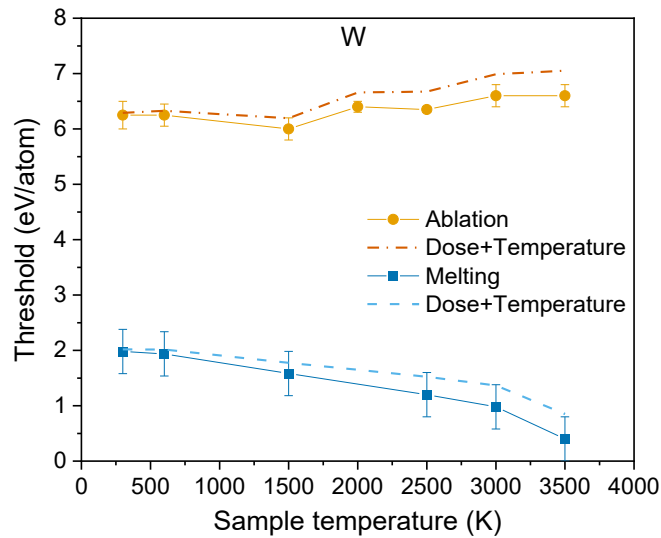

*Figure 5. Damage threshold in tungsten vs. irradiation temperature, simulated within NVE ensemble with XTANT-3 code, with open boundaries along the Z axis. Points are calculated damage threshold, connected by lines to guide the eye. Circles show the ablation threshold, squares stand for the melting threshold. The dashed line is the calculated threshold with added kinetic energy associated with the pre-heating of the material.*

The melting threshold in tungsten is decreasing with the increase of the in-situ temperature of the material. Adding the energy associated with the initial temperature to the threshold does not describe the threshold lowering for the atomic temperatures above ~1500 K. At higher temperatures, the tungsten layer is damaged at lower doses than expected from the sum of the heating and the deposited energy. This is associated with the material expansion of the thin layer simulated, which softens the potential destabilizing the atomic lattice [49]. Thus, we conclude that the hypothesis of the material damage defined solely by the total content of energy in the sample is not always valid. When other than thermal channels contribute to damage (non-thermal or mechanical), not only the deposited energy, but also particular state of the material plays a role in the damage kinetics.

At the highest temperature studied, 3500 K which is slightly above the calculated melting point, the material starts melting at the surfaces at the start of the simulation – before irradiation, whereas the bulk is still crystalline. Thus, the last point in Figure 5 has the error bar going down to zero. The bulk of the sample melts at the dose of ~0.4 eV/atom within a picosecond of the simulation.

However, it is important to emphasize that the calculated threshold only reflects the onset of melting, but not the final state of the material. In simple metals, the recrystallization process may be very efficient, resulting in nearly complete recovery after cooling [53]. Thus, the melting threshold in Figure 5 should serve as an estimate of the transient material damage, not the *post-mortem*, which may be expected to be higher.

We also estimate an ultrafast ablation threshold, which results in irreversible damage (material removal). For that, we performed a separate series of simulations with ultrafast pulses irradiation (10 fs FWHM) and identified the threshold dose similarly to the previous Sections.

In the case of an intense ultrashort pulse, in finite-size metals, another channel of damage may occur: nonthermal ablation [49]. It is the result of an increase in the electronic pressure that drives atoms of the metal to accelerate and be emitted from the surface. In a nano-layer, an increase of the electronic pressure rips the layer in half in the middle, and the two spall fragments separate. The threshold of this process as a function of the atomic temperature is also shown in Figure 5. Since this damage channel – ultrafast ablation or spallation – is caused by the pressure and not the temperature, it is almost insensitive to the starting atomic temperature. A slight increase of the ablation threshold with the increase of the atomic temperature is attributed to the dampening of pressure in melting material if the melting starts before ablation. This is the case for the high atomic temperatures presented here.

### 3.4 Poly(methyl methacrylate) – PMMA

The nature of damage in organic polymers is different from those in elemental solids. Apart from macroscopic damage such as phase transitions, polymers may experience such processes as molecular desorption, formation of local defects, chain scissions, or cross-linking [54,55]. We model poly(methyl methacrylate) (PMMA) damage under XUV/x-ray irradiation following the methodology developed in our previous work [37]. We identify the damage threshold by the appearance of defects within the simulated polymer. It is known that the threshold of damage formation on the surface, such as atomic emission/erosion, in polymers coincides with the damage threshold in the bulk, thus it can be estimated with NVE (microcanonical) ensemble simulations [37].

Following the methodology from Ref. [37], the damage formation was identified by the appearance of defect energy levels in the band structure of the sample. We identify the formation of transient defects, as well as defects stable within the simulation time. The formation of defects indicates the onset of damage.

The results are shown in Figure 6 up to the ceiling temperature of PMMA (~220 °C [56]). The damage threshold slightly reduces with an increase in the *in-situ* atomic temperature, but the absolute value of the reduction is small, only about ~25%. The reason for this seems to be that the defects are formed predominantly *via* the nonthermal channel, and changes in the sample temperature do not influence this process very strongly.

It is interesting to note that adding the energy associated with the pre-heating of PMMA to the damage threshold shows a slight increase (see the dashed line in Figure 6). The stability of PMMA seems to increase – more overall energy (combined pre-heating and irradiation) is required to damage the material. This indicates that pre-heating helps to recover transient defects formed by irradiation.

We can conclude that the *ultrafast* damage processes are not very sensitive to the PMMA target temperature. However, we can expect thermal effects, including possible phase transitions or damage formation, to be taking place in PMMA at longer timescales (tens of ps or ns), which we cannot access with the current simulation. Those processes are expected to be directly dependent on the target temperature, similarly to elemental materials from the previous Sections.

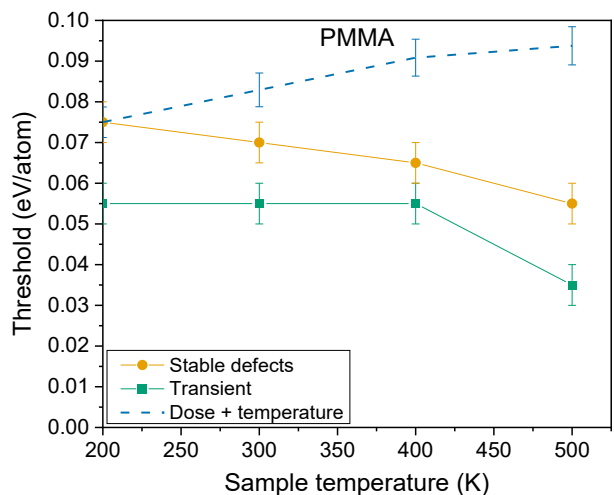

*Figure 6. Damage thresholds in PMMA vs. irradiation temperature, simulated within NPH ensemble with XTANT-3 code. Points are calculated damage threshold, connected by lines to guide the eye. The threshold for transient defects formation is shown with squares, stable defects formation threshold is shown with circles. The dashed line is the calculated threshold with added kinetic energy associated with the pre-heating of the material.*

### 3.5 Implications for prospective experiments

There are in principle two strategies for how experiments can be carried out to uncover the influence of elevated sample temperatures on damage threshold and mechanisms (for a review see Ref. [57] and references therein):

(a) The sample of an investigated material can be heated up by a conventional, e.g., resistive (Joule) heating to a chosen temperature which is controlled by a well-established thermometric technique, and then XUV/x-ray irradiation is performed under a standard protocol for the determination of the damage threshold.

(b) Spatio-temporal distribution of temperatures in the repeatedly irradiated sample is registered experimentally or estimated theoretically to be considered during an investigation of an interaction of the subsequent incoming pulse and its consequences.

For such experiments, the material of the first choice should be diamond. It follows from both the above-reported computer simulations (Fig. 2; even for pre-heating at a temperature below 1000°C the threshold lowering caused by the pre-heating seems to be considerable) and the great technical importance of the diamond. Diamond materials are widely used in optical elements and detectors at x-ray laser facilities [12,13,15] and a degree of graphitization can be revealed by Raman spectroscopy [58].

## 4. Conclusion and outlook

We presented a theoretical study of the damage threshold in a few materials under ultrafast XUV/x-ray irradiation. The dependence of the damage threshold on the pre-heating temperature was identified in diamond, silicon, tungsten, and PMMA. Each material and damage channel exhibited its own dependence on the *in-situ* atomic temperature, showing that the damage cannot be unambiguously defined by the total energy content in sample (delivered *via* preheating and irradiation). The thermal damage (induced by the increase of atomic temperature *via* electron-phonon coupling) is typically reduced with an increase in the pre-heating temperature, whereas nonthermal damage channels (induced by changes of the interatomic potential due to excitation of electrons) may change in either direction – increase or decrease.

The thresholds of nonthermal graphitization in diamond and nonthermal melting in silicon decrease with the increase of the atomic temperature. Nonthermal defect formation in PMMA is also slightly reduced by the absolute value. The nonthermal ablation of tungsten, in contrast, increases with the increase of the pre-heating temperature, which is associated with pressure dissipation.

These results have implications for studies of multi-shot damage of materials, exposed to repeated irradiations. Materials that do not have time to cool down before the arrival of a subsequent pulse may experience damage at doses different from those expected from the energy content of the residual temperature and the deposited dose. Therefore, it is very important to account for the dependence of the damage threshold on the atomic temperature of the material.

## 5. Conflict of Interest

The authors declare no conflict of interests, financial or otherwise.

## 6. Data Availability Statement

The data that support the findings of this study are available from the corresponding author upon reasonable request.

## 7. Acknowledgments

Computational resources were supplied by the project "e-Infrastruktura CZ" (e-INFRA LM2018140) provided within the program Projects of Large Research, Development and Innovations Infrastructures. NM gratefully acknowledges financial support from the Czech Ministry of Education, Youth and Sports (grant No. LM2018114). ZK greatly appreciates the financial support provided by the Czech Ministry of Education, Sports and Youths (Mobility project No. CZ.02.2.69/0.0/0.0/18_053/0016627) making her research stay at DESY in Hamburg possible. JCh greatly appreciates the financial support provided by the Czech Science Foundation (grant No. 20-08452S). This research team is also supported within the European Union's Horizon 2020 research and innovation programme under grant agreement no. 871124 Laserlab-Europe. This work benefited from networking activities carried out within the EU funded COST Action CA17126 (TUMIEE) and represents a contribution to it.